\documentclass[]{IEEEtran}
\usepackage{amsfonts,amsmath,amssymb,url,graphicx,epsfig}

\usepackage[english]{babel}
\usepackage[ruled,vlined]{algorithm2e}
\usepackage{algpseudocode}
\usepackage{mathrsfs}
\usepackage{pifont}
\usepackage{array}
\usepackage{tabularx,ragged2e,booktabs,caption}
\usepackage{multirow}
\graphicspath{{./figs/}}
\usepackage{etoolbox}
\usepackage{blindtext}
\newcommand{\subparagraph}{}
\usepackage{titlesec}
\usepackage{color}
\usepackage{flushend}
\usepackage{amssymb}
\usepackage{multirow}
\usepackage{subfig} 
\usepackage{caption}
\usepackage[
backend=biber,
style=ieee,
sorting=none
]{biblatex}
\usepackage{svg}
\graphicspath{{images/}}

\makeatletter
\let\old@ps@headings\ps@headings
\let\old@ps@IEEEtitlepagestyle\ps@IEEEtitlepagestyle
\def\confheader#1{%
\def\ps@headings{%
\old@ps@headings%
\def\@oddhead{\strut\hfill#1\hfill\strut}%
\def\@evenhead{\strut\hfill#1\hfill\strut}%
}%
\def\ps@IEEEtitlepagestyle{%
\old@ps@IEEEtitlepagestyle%
\def\@oddhead{\strut\hfill#1\hfill\strut}%
\def\@evenhead{\strut\hfill#1\hfill\strut}%
}%
\ps@headings%
}
\makeatother
\IEEEoverridecommandlockouts 
\begin{document}
\title{Partially Permuted Multi-Trellis Belief Propagation for Polar Codes}
\author{ \IEEEauthorblockN{Vismika Ranasinghe, Nandana~Rajatheva,  and Matti Latva-aho}

\IEEEauthorblockA{Centre for Wireless Communications,~~University of Oulu, Finland\\
E-mail: vismika.maduka@oulu.fi, nandana.rajatheva@oulu.fi, matti.latva-aho@oulu.fi}
}
\maketitle
\thispagestyle{empty}

\setlength{\textfloatsep}{8pt plus 1.0pt minus 2.0pt}
\titlespacing{\section}{0pc}{0.55pc}{0.55pc}
\titlespacing{\subsection}{0pc}{0.25pc}{0.25pc}
\begin{abstract}
Belief propagation (BP) is an iterative decoding algorithm for polar codes which can be parallelized effectively to achieve higher throughput. However, because of the presence of error floor due to cycles and stopping sets in the factor graph, the performance of the BP decoder is far from the performance of state of the art cyclic redundancy check (CRC) aided successive cancellation list (CA-SCL) decoders. It has been shown that successive BP decoding on multiple permuted factor graphs, which is called the multi-trellis BP decoder, can improve the error performance. However, when permuting the entire factor graph, since the decoder dismisses the information from the previous permutation, the number of iterations required is significantly larger than that of the standard BP decoder. In this work, we propose a new variant of the multi-trellis BP decoder which permutes only a subgraph of the original factor graph. This enables the decoder to retain information of variable nodes in the subgraphs, which are not permuted, reducing the required number of iterations needed in-between the permutations. As a result, the proposed decoder can perform permutations more frequently, hence being more effective in mitigating the effect of cycles which cause oscillation errors. Experimental results show that for a polar code with block length 1024 and rate 0.5 the error performance gain of the proposed decoder at the frame error rate of $10^{-6}$ is 0.25 dB compared to multi-trellis decoder based on full permutations. This performance gain is achieved along with reduced latency in terms of the number of iterations.   

\end{abstract}

\section{Introduction}


Polar codes, the only known capacity achieving error correcting code \cite{arikan} has been selected  as the channel code of the control channel of current 5\textsuperscript{th} generation mobile communication standard by the 3GPP \cite{tsgR1}. Hence, decoder implementation for polar codes has become more of a practical challenge. There are two methods to decode  polar codes, one is successive cancellation (SC) \cite{arikan} and its derivatives which are more serial in nature, while the other method is belief propagation (BP) proposed in \cite{arikan2010polar}  which can be parallelized easily making it an ideal candidate for high throughput applications. Since the state of the art derivatives of SC algorithm like successive cancellation list (SCL) decoding and cyclic redundancy check(CRC) aided SCL (CA-SCL) \cite{talvardy}  outperform BP decoding by a significant margin, it has become an active area of research to make the performance of BP decoders comparable to that of state of the art SC algorithms.

BP is an iterative message passing algorithm on a factor graph which gained its popularity in decoding LDPC codes. Being an iterative decoding algorithm BP decoder for polar codes suffers from error floor in high SNR regime and a number of investigations were carried out in understanding and mitigating the error floor  \cite{clippingeffect}-\cite{stopping_sets}.  In the high SNR regime, the main contributors  to the error floor are  false convergences and oscillations caused by cycles and stopping sets in the polar code factor graph \cite{errorpattern}. Shorter the cycles and smaller the stopping sets, their effects on error performance become  more significant \cite{stopping_sets}. Concatenation with high rate error detection schemes like CRC can alleviate the effect of oscillation errors to some extent \cite{errorpattern}. However, this reduces the effective code rate. Work in \cite{postprocesing} shows that in addition to CRC concatenation, detection of false convergences and oscillations proceeded by  post-processing improves the performance in high SNR regime at the cost of decoding complexity. In \cite{clippingeffect} to mitigate the error floor the multi-trellis BP decoder is suggested where over complete representation \cite{polarUrbanke} of the polar code is utilized by the decoder with the aid of CRC to successively perform belief propagation on different permutations of the original factor graph \cite{permuteFactor}. Since the implementational complexity of the decoder linearly increases with the number of factor graphs used, decoding on all permutations becomes impractical, hence either a limited number of random permutations of the original factor graph or cyclic shifts of the original factor graph are used. Furthermore, \cite{permuteFactor} put forward the idea of partitioned permuted factor graphs similar to the partitioned successive cancellation list (PSCL) decoder proposed in \cite{partitionedSCL} to reduce complexity at the expense of performance. Further improvements to the multi-trellis BP decoder can be made by carefully choosing the permutations of the original factor graph based on the numerically evaluated error correcting performance \cite{permuteWarrenJ}. Authors in \cite{permuteList} propose the use of independent BP decoders to perform the belief propagation on different permutations of the original factor graph concurrently generating a list of possible transmitted codewords. Here the one closest to received vector in euclidean distance is chosen, reducing the latency by avoiding expensive--in terms of latency--successive decoding on different permutations. In \cite{clippingeffect} virtual noise is injected to the decoder to mitigate the effect of clipping on error floor.  

\begin{figure}[htb]
  \centering
  \includegraphics[height =5cm, width =8cm]{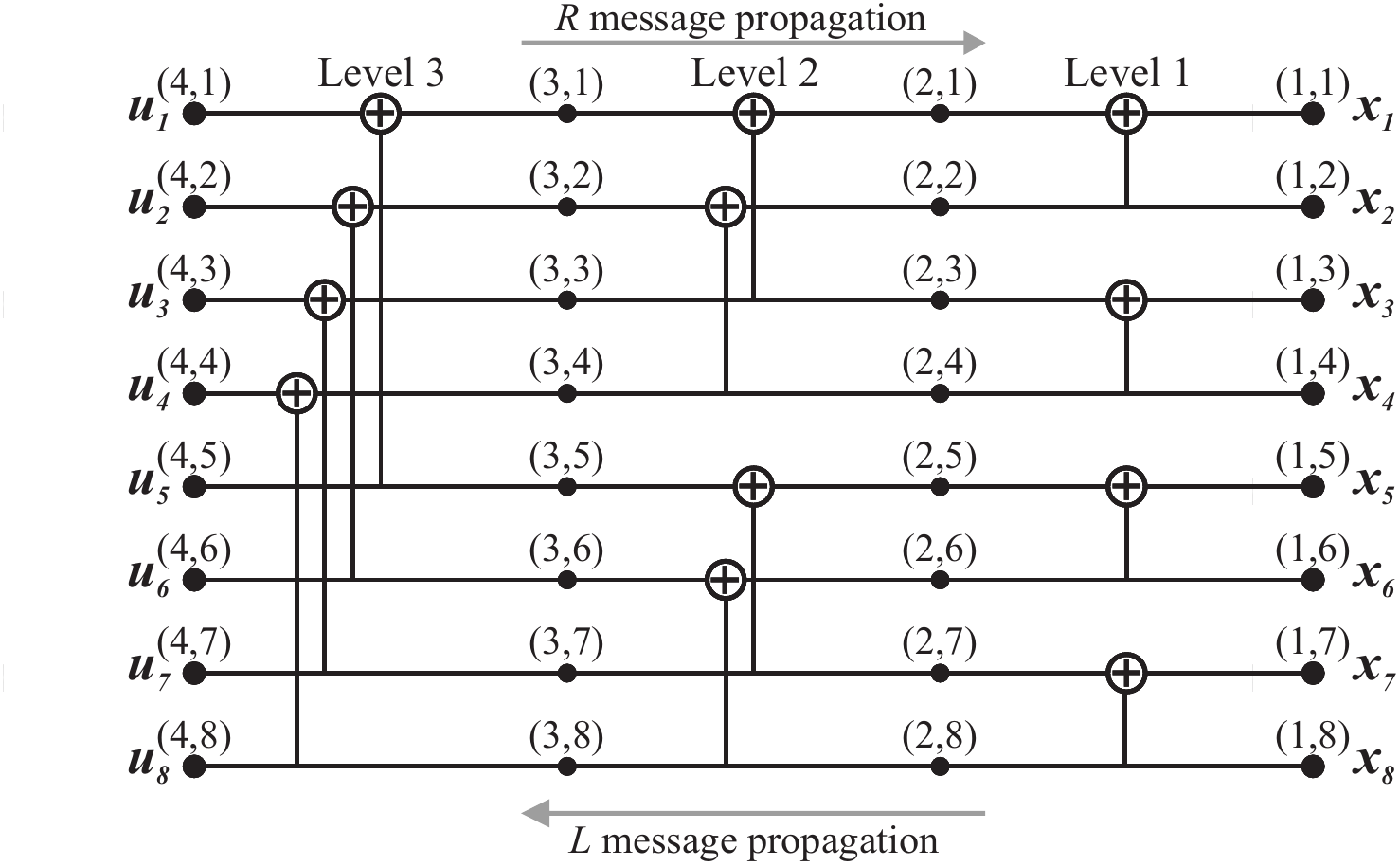}
  \caption{Factor graph of N=8 polar code}
  \label{factorgraph}
\end{figure}

In this work, we propose a variant of the multi-trellis BP decoder based on partially permuted factor graphs. We show that by only permuting a subgraph of the original factor graph, the decoder is able to retain the information of the variable nodes in the subgraphs that are not permuted. This results in more frequent permutations within the given number of iterations. Our simulation results show that for a polar code with block length 1024 and rate 0.5 the gain in error performance is around 0.25 dB at frame error rate of $10^{-6}$ compared to the multi trellis BP decoder \cite{permuteFactor}. This improvement in error correcting performance is achieved along with reduced latency.

The remainder of paper is as follows. Section II  gives a short introduction to polar codes and belief propagation. Further, it briefly explains the classification of BP decoding errors and factors that affect the distribution of errors. Section III and Section IV present the proposed decoder and experimental results respectively. Finally, a conclusion is provided in Section V

\section{Background}
\subsection{polar codes}

\begin{figure}[htb]
  \centering
  \includegraphics[height =2.5cm, width =4cm]{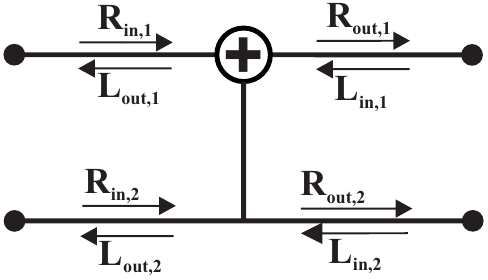}
  \caption{Processing Element}
  \label{process_element}
\end{figure}

Based on the theoretical concept of channel polarization, polar code is the only type of channel code proven to achieve capacity under SC decoding for infinite block lengths \cite{arikan}. $N$ identical channels are combined recursively using a $2\times 2$  kernel matrix to create $N$ synthesized channels which exhibit the polarization effect. Due to this polarization effect each synthesized channel is converged to either a pure noiseless channel or a completely noisy channel. As a result, the information bits are only transmitted on pure noiseless channels while transmitting a known set of bits which are called frozen bits on noisy channels.  

A number of algorithms are available on how to perform the polar code construction which is selecting the set of  noiseless channels for the information set $\mathbb{A}$ and completely noisy channels for the frozen set  $\mathbb{A}^c$. In this paper, we use the polar code construction based on Arikan's Bhattacharyya bounds \cite{arikan} of the bit channels designed at $E_b/N_0 = 0$ dB. Encoding of the polar codeword of length $N = 2^n$  is based on the polarization matrix of $N\!\!\times \!\! N$ dimension  given by

\begin{equation}
\mathbf{G}_N = \mathbf{F}^{\otimes n } , \;\;\;\;\;   \mathbf{F} = \begin{bmatrix}
1&0\\
1&1\\
\end{bmatrix}
\end{equation}

 where $\mathbf{F}^{\otimes n }$ denotes the $n^{th}$ Kronecker power of $\mathbf{F}$. The encoded codeword $\mathbf{x}$ can be obtained by $\mathbf{x} = \mathbf{uG}_N $ where $\mathbf{u}$ is the input to the decoder with $k$ information bits and $N-k$ frozen bits. 
 \subsection{Belief propagation}
 
 Belief propagation is an iterative algorithm which determines the information bits by passing messages, which are typically log-likelihood ratios (LLR), on a factor graph based on the polarization matrix $\mathbf{G}_N$. As illustrated in Figure \ref{factorgraph}, a factor graph consists of variable nodes identified by tuples $(i,j), 1 \leq i \leq n+1, 1\leq j \leq N$ and $N/2$ processing elements at each level. $\mathbf{R}_{i,j}$ messages (from left to right) and $\mathbf{L}_{i,j}$ messages (from right to left) are the two types of messages that are exchanged between variable nodes where tuple $(i,j)$ denotes the variable node from which the message is originated. Variable nodes $(n+1,j)$ hold the priori information ($\mathbf{R}_{n+1,j}$) available to decoder based on the code construction, and therefore initialized to 0 and $\inf$ for non-frozen and frozen bits respectively. $(1,j)$ nodes are initialized with the LLR from channel output ($\mathbf{L}_{1,j}$). Then the messages are iteratively updated by processing element (PE) as shown in Figure \ref{process_element} according to the following equations:
\begin{align*}
    R_{out,1} &= f(R_{in,1},L_{in,2} + R_{in,2}) \\
    R_{out,2} &= f(R_{in,1},L_{in,1}) + R_{in,2} \\
    L_{out,1} &= f(L_{in,1},L_{in,2} + R_{in,2}) \\
    L_{out,2} &= f(R_{in,1},L_{in,1}) + L_{in,2} \\
\end{align*}
   
where $f(x,y) = x \boxplus y$, known as $boxplus$ operator \cite{turbo}. $Boxplus$ operator can be approximated by $sign(x)sign(y)min(x,y)$ reducing the implementational complexity of the decoder at a cost of small performance degradation. In the decoding process after the maximum number of iterations ($N_{it,max}$) is reached or early stopping criterion is met, information bits are estimated to be 0 or 1 if $\mathbf{R}_{n+1,1} + \mathbf{L}_{n+1,1} \geq 0$ or otherwise respectively.

\subsection{Error classification and distribution}

According to the error classification done in \cite{errorpattern} there are three types of errors that contribute to the BP decoding errors in different proportions depending on $E_b/N_0$.

\begin{itemize}
  \item \textit{False converged}: if the decoder converges to an incorrect codeword, it is classified as false converged error and caused by decoder falling into local minima.
  \item \textit{Oscillation}: if the decoder oscillates between two codewords over iterations, it is called an oscillation error and caused by few incorrect decisions reinforcing themselves through cycles in the factor graph.
  \item \textit{Unconverged}: if the decoder fails to converge or does not result in an oscillation, it is known as an unconverged error and caused by noisy input to the decoder.
\end{itemize}

Looking at the statistical breakdown of BP decoding errors in \cite{errorpattern}, it is evident that in high SNR regime decoder errors are mostly dominated by false convergences or oscillations depending on the code design parameters. In high rate codes, decoding errors  are predominately caused by false convergences which is a result of more crowded codeword space, hence smaller minimum stopping sets. Codes with large block lengths suffer mainly from oscillation errors due to the presence of more cycles in the factor graph \cite{stopping_sets}\cite{errorpattern}.

\subsection{Permuted factor graphs}
For a polar code of length $2^n$, as outlined in \cite{polarUrbanke} there exists $n!$ permutations of the factor graph which are valid for $\mathbf{x} = \mathbf{uG}_N $, hence BP decoding can be performed in any of such permutations. The decoding of polar codes on permuted factor graphs is referred to as ``multi-trellis BP decoding"  and is proposed in \cite{clippingeffect} to mitigate the error floor caused by LLR-clipping. In multi-trellis BP decoding \cite{permuteFactor}, BP iterations are performed on a random permutation of the original factor graph until the number of iterations per each permutation ($N_{it,reset}$) is reached or the stopping criterion is satisfied. If stopping criterion is not satisfied within $N_{it,reset}$, priori information $\mathbf{R}_{n+1,j}$ and channel output $\mathbf{L}_{1,j}$ are passed on to a new permuted factor graph until predefined number of factor graphs ($q_{max}$) are tried \cite{permuteFactor}. 

Permuting the factor graph alters the stopping sets and cycles on it. Hence, resulting in different error performances under the same noise realization. In \cite{permuteFactor} it is shown that if a codeword can be successfully decoded by an SCL decoder which achieves the ML bound, there exists one factor graph representation where the BP decoder will converge to the correct codeword under a carefully chosen stopping criterion.

\section{Multi-trellis BP decoding on partially permuted factor graphs}

 In BP decoding extrinsic information is exchanged between variable nodes in an iterative manner until convergence is achieved. The cycles in factor graph degrade the uncorrelatedness of the information exchanged, and thus adversely affecting the error performance. The shorter the cycles in the factor graph, the exchanged information become correlated in a smaller number of iterations. The shortest cycle contained in the factor graph is called \textit{Girth} and extrinsic information exchanged at every variable node remains uncorrelated until the number of iterations reaches half the length of the girth. Furthermore, the girth of the original factor graph of any block length is constant and equals to 12 \cite{stopping_sets}. It has been shown that for LDPC codes the error performance gain diminishes as we increase the maximum number of iteration per codeword ($N_{it,max}$) \cite{ldpc_perf}. This saturating behavior of gain in error performance is a result of cycles in the factor graph which cause the exchanged information between variable nodes to become correlated. Hence, in polar codes the saturating behavior can be alleviated by permuting the factor graph. This permutation changes the cycles in the factor graph by deleting the existing cycles and creating cycles among a new set of variable nodes.

In multi-trellis BP decoding, if the stopping criterion is not satisfied after the maximum number of iterations ($N_{it,reset}$) on the original factor graph, belief propagation is performed on a new permutation of the original factor graph. Hence, the decoder permutes the factor graph in every $N_{it,reset}$ iterations until the maximum number of factor graph permutations ($q_{max}$) is reached or the stopping criterion is met. Since each permutation of the original factor graph has different cycles, increasing $N_{it,max}$ along with $q_{max}$ enable further performance gains \cite{permuteFactor} compared to that of the standard BP decoder. Here, $N_{it,max} =  N_{it,reset} \times q_{max}$.

\begin{figure}[htb]
\centering
\subfloat[Conventional factor graph]{\label{ppermute1_cycle}{\includegraphics[height =4cm, width =4.2cm]{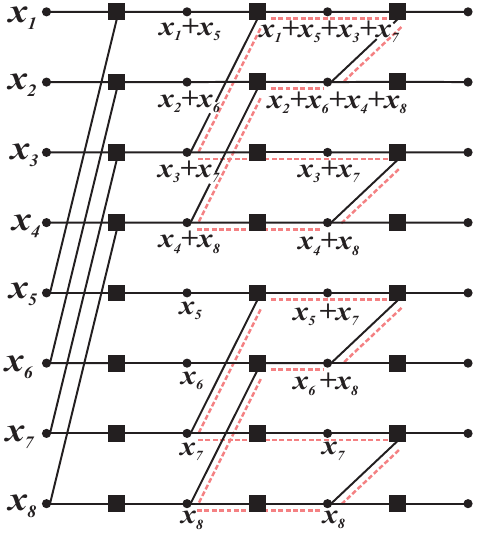}}}\hfill
\subfloat[Permuted factor graph]{\label{ppermute2_cycle}{\includegraphics[height =4cm, width =4.2cm]{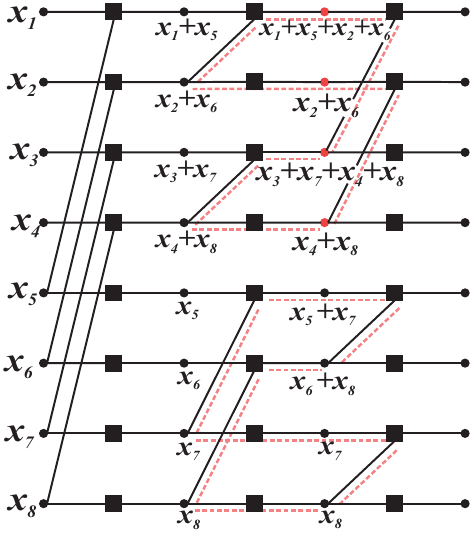}}}
\caption{Variable nodes and cycles in a original factor graph and a partially permuted factor graph with $\rho_{range} = 2$ and $\rho_{level} = 1$} \label{fig:ppermute_cycle}
\end{figure}

A key drawback of multi-trellis BP decoding is discarding the intermediate variable node values when permuting the factor graph. Only the priori information $\mathbf{R}_{n+1,j}$ and LLR from the channel output $\mathbf{L}_{1,j}$ are preserved on the factor graph after the permutation. However, in the event of permutation, only the variable nodes in between the permuted levels get modified (Figure \ref{fig:ppermute_cycle}) which means LLR values of other variable nodes can be utilized by belief propagation after the permutation. Utilizing LLR values of intermediate variable nodes from the previous permutation on the next permutation enables the decoder to have smaller $N_{it,reset}$ on that permutation. As a result, in this work, we propose an improved variant of the multi-trellis decoder which is based on partial permutations of factor graph while retaining the information of the variable nodes which do not change. Permuting only a subgraph enables the decoder to realize new permutations with minimal modifications to intermediate variable nodes resulting in minimal loss of information between the permutations. This results in more frequent permutations with smaller $N_{it,reset}$ which can be effective against the effects of cycles. Henceforth, the abbreviations BP, FP-BP (fully permuted belief propagation), and PP-BP (partially permuted belief propagation) are used for the standard BP decoder, multi-trellis decoder proposed in \cite{permuteFactor}, and the decoder proposed in this work respectively.

PP-BP requires more configuration parameters than that of FP-BP and BP as given in Algorithm \ref{multitrellis}. $P_{range}, 2 \leq P_{range} \leq n$ is the maximum number of levels in the factor graph that can be permuted in a single permutation while $P_{level}, 1 \leq P_{level} \leq n-1$ determines the maximum level (Figure \ref{factorgraph}) the rightmost level of any given permutation can be. Similar to the conventional BP decoder, PP-BP performs belief propagation iteratively on the original factor graph until $N_{it,reset}$ is reached or stopping criterion is met. At the beginning of the decoding process $N_{it,reset}$ is initialized based on the block length $N$ and $N_{it,max}$.  If the $N$ is large, a sufficient number of iterations should be performed before permuting the factor graphs to enable the maximum exchange of uncorrelated extrinsic information between the variable nodes meanwhile making sure within $N_{it,max}$ sufficient number of permutations are performed to alter the cycles. In case the stopping criterion is not met after $N_{it,reset}$ iterations, a subgraph of the previous factor graph is permuted. Number of stages to be permuted in subgraph ($\rho_{range}$) is selected randomly from the set \{$z \in \mathbb{Z}: 2 \leq z \leq P_{range}$\} and the stage at which the permutation begins ($\rho_{level}$) is $\min(x,n-\rho_{range}+1)$ where $x$ is randomly chosen from the set \{$z \in \mathbb{Z}: 1 \leq z \leq P_{level}$\}. Since the variable $\rho_{level}$ is the rightmost level of the permutation, $\rho_{level}$ must satisfy the condition $\rho_{level} \leq n-\rho_{range}+1$. Once the $\rho_{range}$ and $\rho_{level}$ values are chosen, from the available number of subgraphs which is $\frac{N}{2^{\rho_{level}+\rho_{range}-1}}$, a single subgraph is chosen randomly and the levels from $\rho_{level}$ to $\rho_{level} + \rho_{range}-1$ are permuted. Figure \ref{ppermute2_cycle} illustrates a partially permuted factor graph for polar code of $N = 8$ where the uppermost subgraph with $\rho_{range} = 2$ and $\rho_{level} = 1$ is permuted. After the permutation, the LLR values of the modified variable nodes are set to 0 as they are no longer valid. Then the number of variable nodes modified ($N_{zero}$), which is determined by $N_{zero} = 2^{\rho_{level}+\rho_{range}-1}\times(\rho_{range}-1)$, is calculated and for the next permutation $N_{it,reset}$ is calculated based on $N_{zero}$. In this work $N_{it,reset}$ is calculated as

\begin{equation}
    N_{it,reset} = \max(N_{min}, \lfloor \frac{N_{zero}}{D} \rfloor)
    \label{eq:resetvalue}
\end{equation}

where the parameters $N_{min}$ and $D$ are decoder specific configuration parameters.

\begin{algorithm}[htb]
\SetAlgoLined
\DontPrintSemicolon
\SetKwInput{Input}{Input} 
\SetKwInput{Output}{Output} 
\Input
  {\begin{itemize}
  \setlength\itemsep{0em}
   \item[] {\makebox[2cm][l]{$\mathbf{L}_{1,j}$}      \% LLR channel output}
   \item[] {\makebox[2cm][l]{$\mathbb{A}$}  \% information set}
   \item[] {\makebox[2cm][l]{$P_{level}$} \% max. level the rightmost level of the subgraph can be}
    \item[] {\makebox[2cm][l]{$P_{range}$} \% max. no. of stages to be permuted}
   \item[]  {\makebox[2cm][l]{$N_{it,max}$} \% max.no.of iterations}
   \item[]  {\makebox[2cm][l]{$stopI\!D$}  \% stopping criterion}
   \item[]  {\makebox[2cm][l]{$D$}  \% decoder specific parameters}
   \item[]  {\makebox[2cm][l]{$N_{min}$}}
   \end{itemize}
  }

  \Output
  {\begin{itemize}
  \setlength\itemsep{0em}
      \item [] {\makebox[2cm][l]{$\mathbf{\hat{u}}$} \% estimated codeword}
  \end{itemize}
  }

$N \longleftarrow  length(\mathbf{L}_{1,j})$\;
$n \longleftarrow \text{log}_2N$\;
$ (\mathbf{L},\mathbf{R}) \longleftarrow \mathbf{initializeLandR(\mathbf{L}_{1,j},\mathbb{A}})$\; 
$N_{it,reset} \longleftarrow \mathbf{initializeIterationResetValue}(N)$\;

\For{$iter\leftarrow 1$ \KwTo $N_{it,max}$}{
$ (\mathbf{L},\mathbf{R}) \leftarrow \mathbf{oneBPiteration(N, n, \mathbf{L}, \mathbf{R}},schedule)$\;

\If{$checkStopCondition(\mathbf{L},\mathbf{R},stopID)$}
{
$\mathbf{\hat{u}} = hardDecision(\mathbf{L}_{n+1,j}, \mathbf{R}_{n+1,j})$\;
$\mathbf{return \; \hat{u}}$
}
\If{$iter == N_{it,reset}$}
{
$\rho_{range} = randomRangeValue(P_{range})$\;
$\rho_{level} = randomLevelValue(P_{level}, \rho_{range})$\;
$schedule \leftarrow partialPermute(\rho_{range}, \rho_{level})$\;
$ (\mathbf{L},\mathbf{R}) \leftarrow \mathbf{resetVariableNodes}(N,n,\mathbf{L}, \mathbf{R}, schedule)$\;
$N_{zero} = 2^{\rho_{level}+\rho_{range}-1}\times(\rho_{range}-1)$\;
$N_{it,reset} = \max(N_{min}, \lfloor \frac{N_{zero}}{D} \rfloor)$\;
$N_{it,reset} = iter+ N_{it, reset}$
}
}

$\mathbf{\hat{u}} = hardDecision(\mathbf{L}_{n+1,j}, \mathbf{R}_{n+1,j})$\;

 \caption{Partially Permuted Multi-Trellis BP decoding}
 \label{multitrellis}
\end{algorithm}

\section{Experimental Results and Analysis}

In this section, we evaluate the performance of proposed partially permuted belief propagation (PP-BP) decoder and compare it against the belief propagation (BP), fully permuted belief propagation (FP-BP) \cite{permuteFactor}, and noise aided belief propagation (NA-BP) \cite{clippingeffect} decoders. Furthermore, we compare it against the state of the art CRC aided successive cancellation list (CA-SCL) decoder with list size $L=16$. For this simulation, a system model with additive white Gaussian noise (AWGN) channel and quadrature phase shift keying (QPSK) modulation is used. Moreover, the boxplus operator in belief propagation is approximated with nonscaled min-sum approximation. All the variants of BP decoder and the CA-SCL decoder in this section are aided by 24-bit CRC. Here, in variants of BP decoder in order to allow the soft output of the decoder to be stable, the stopping criterion is not applied in first few iterations. Even though the CRC bits are included in rate $R$,  in the simulations energy per symbol is calculated using the number of information bits excluding the CRC bits. Since the error distribution of BP decoding \cite{errorpattern} depends on the number of information bits including the CRC bits, in the code rate calculation CRC bits are included.

\begin{table}[]
\renewcommand{\arraystretch}{1.25}
\resizebox{\columnwidth}{!}{
\begin{tabular}{|l|l|l|}
\hline
$N_{it,max}$       & \multicolumn{1}{c|}{Simulation} & \multicolumn{1}{c|}{Parameters} \\ \hline
\multirow{3}{*}{$2\times10^4$} & FP-BP[N=512/1024 R=0.5/0.65]&  $q_{max}\! \!=\! \!100$; $N_{it,reset}\! \! =\! \! 200$                               \\ \cline{2-3} 
                  & FP-BP [N=1024; R=0.5] & $q_{max} \! \!= \! \!200$; $N_{it,reset}\! \!=\! \!100$ \\ \cline{2-3} 
                  & NA-BP [N=1024 R=0.5/0.65] &  $\sigma^2_{noise}\! \!=\! \!0.36$ \cite{clippingeffect}
                  \\ \cline{2-3} 
                  & PP-BP [N=1024 R=0.5/0.65]   & \begin{tabular}[c]{@{}l@{}}$P_{range}\! \! = \! \!10$; $P_{level}\! \!=\! \!9$;\\ $D\! \!=\! \! 100$; $N_{min}\! \!=\! \!15$\end{tabular}
                  \\ \cline{2-3} 
                  & PP-BP [N=512 R=0.5] &  \begin{tabular}[c]{@{}l@{}}$P_{range}\! \!=\! \!9$; $P_{level}\! \!=\! \!8$;\\ $D\! \!=\! \!50$; $N_{min}\! \!=\! \!15$\end{tabular}
                  \\ \hline
\multirow{2}{*}{$2\times10^2$}  & NA-BP [N=1024; R=0.5/0.65] &  $\sigma^2_{noise}\! \! = \! \!0.36$ \cite{clippingeffect}
                  \\ \cline{2-3} 
                 & PP-BP [N=1024 R=0.5/0.65]   & \begin{tabular}[c]{@{}l@{}}$P_{range}\! \!=\! \!2$; $P_{level}\! \!=\! \!6$;\\ $D\! \!=\! \!8$; $N_{min}\! \!=\! \!4$\end{tabular}
             \\ \hline
             
\end{tabular}
}
\caption{Decoder specific simulation parameters for FP-BP, NA-BP, and PP-BP}
\label{table:simparam}
\end{table}

Decoder specific parameters for each simulation is tabulated in Table \ref{table:simparam}. When $N_{it,max} \!\!=\!\! 2\times10^4$ and $N\!=\!1024$ and $512$, for the PP-BP the parameters $P_{range}$ and $P_{level}$ are set to $n$ and $n-1$ respectively, to their maximum allowed values, to enable the decoder to realize largest possible subgraph permutations. In this case, $D$ is chosen such that when the entire factor graph is permuted $\frac{N_{zero}}{D}$, where $N_{zero} \!=\! N\!\times\! (n-1)$, would be approximately equal to the number of iterations ($N_{it,max}$) required to observe the saturating behavior in the BP decoder for block length $N$. In the simulations, it was observed that for both $N=512$ and $1024$ there is no significant gains in performance when $N_{it,max}$ is increased beyond 100. Hence, using the above heuristic $D \approx \frac{N \! \times\! (n-1)}{100}$. To make the results comparable, for FP-BP simulations are carried out with $N_{it,reset}\!=\! 100$ in addition to simulations with $N_{it,reset} \!=\! 200$ similar to the work in \cite{permuteFactor}. For PP-BP when $N_{it,max}$ is limited, $P_{range}$ and $P_{level}$ have to be small to limit the size of the subgraph permuted in each permutation. This will enable the decoder to realize more frequent permutations. In this work $P_{range}$ and $P_{level}$, when  $N_{it,max} = 200$, are found empirically with the aid of above heuristic.

\begin{figure}[htbp]
  \centering
  \includegraphics[width=8.5cm,height = 16cm]{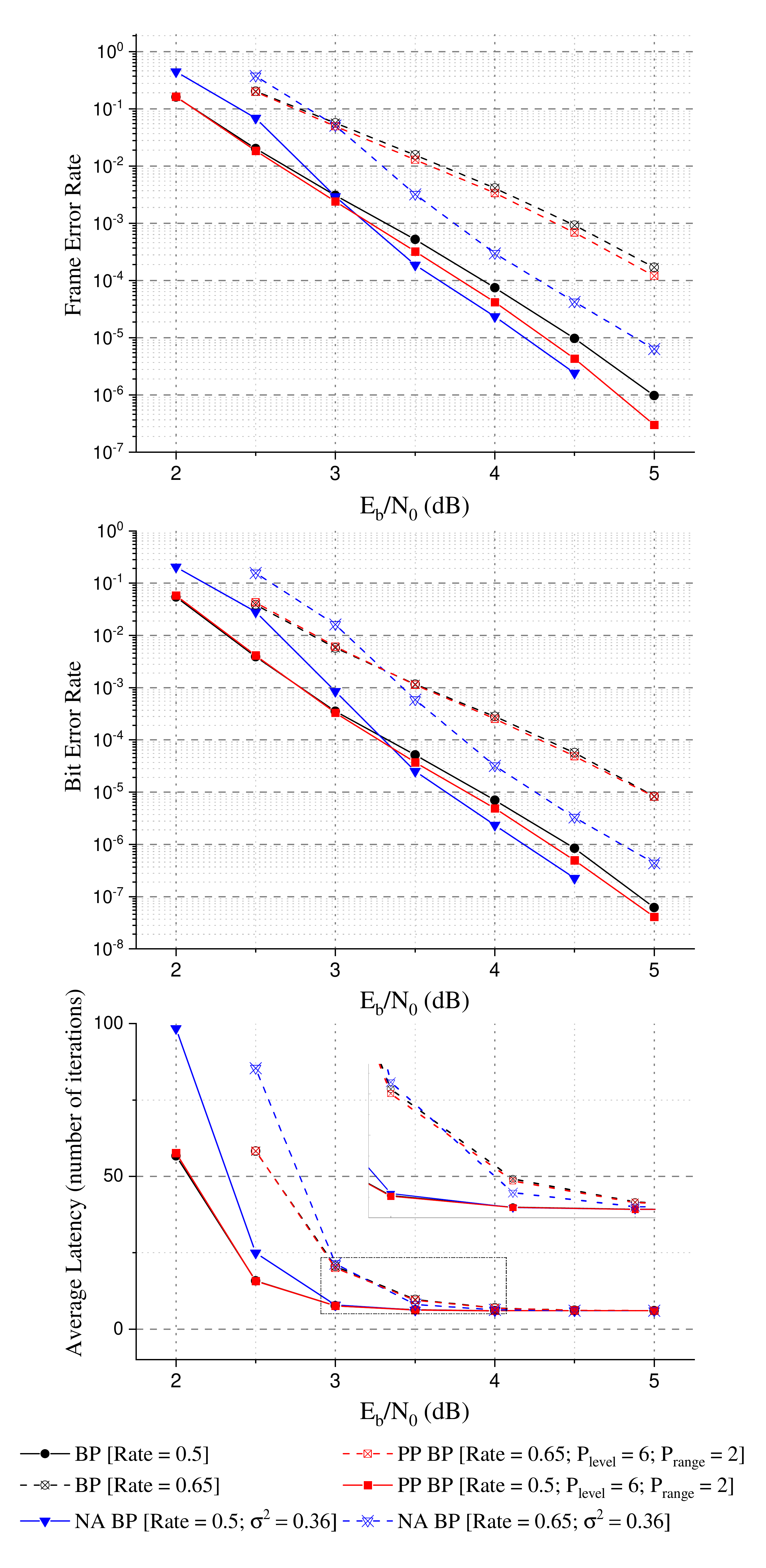}
  \caption{BER, FER and average latency comparison between BP,  NA-BP and PP-BP for $N \!=\! 1024$ polar code with $N_{it,max} \!=\! 200$ and 24-bit CRC}
  \label{200_perf}
\end{figure}

Figure \ref{200_perf} presents a performance comparison between BP, NA-BP, and PP-BP for a polar code with $N \!=\!1024$, $R \!=\! 0\text{.}5/0\text{.}65$, and $N_{it,max}\!=\!200$. It can be observed that NA-BP is outperformed in terms of frame error rate (FER) by both BP and PP-BP in low SNR regime ($E_b/N_0 \leq 3dB$) for both $R\!=\!0\text{.}5$ and $0\text{.}65$. In low SNR regime, since the majority of BP errors are unconverged errors, adding noise to decoder input makes it difficult for the decoder to achieve convergence as it makes the effective channel noisier. For $R\!=\!0.5$, in high SNR regime ($E_b/N_0 \geq 4dB$) both NA-BP and PP-BP outperform the BP by about $0.3dB$ and $0.2dB$ at $10^{-5}$ in terms of FER. Furthermore, when $R\!=\!0\text{.}65$, it can be observed that the NA-BP outperforms PP-BP and BP by about $0.7dB$ at $10^{-3}$ in terms of FER. This significant performance gain compared to $R\!=\!0\text{.}5$ case is a result of the difference in error distribution in high SNR regime when $R \leq 0\text{.}5$ and $R > 0\text{.}5$. When $R \leq 0\text{.}5$, in high SNR regime, a majority of the errors occurred are oscillation errors, while when $R > 0\text{.}5$, the error distribution is dominated by converged errors \cite{errorpattern}. Hence, it can be concluded that for converged errors noise injection is more effective than partial permutations. Comparing the latency in terms of average number of iterations, it can be observed that NA-BP is outperformed by both BP and PP-BP in low SNR regime. Since the injection of noise makes the convergence difficult for the decoder, it causes the decoder to spend more iterations increasing the latency. Furthermore, as the performance of BP and PP-BP is similar in terms of latency, it can be concluded that the partial permutations do not hinder the ability of the decoder to converge to a codeword, when $N_{it,max}$ is limited.

\begin{figure}[htb]
  \centering
  \includegraphics[width =8.5cm]{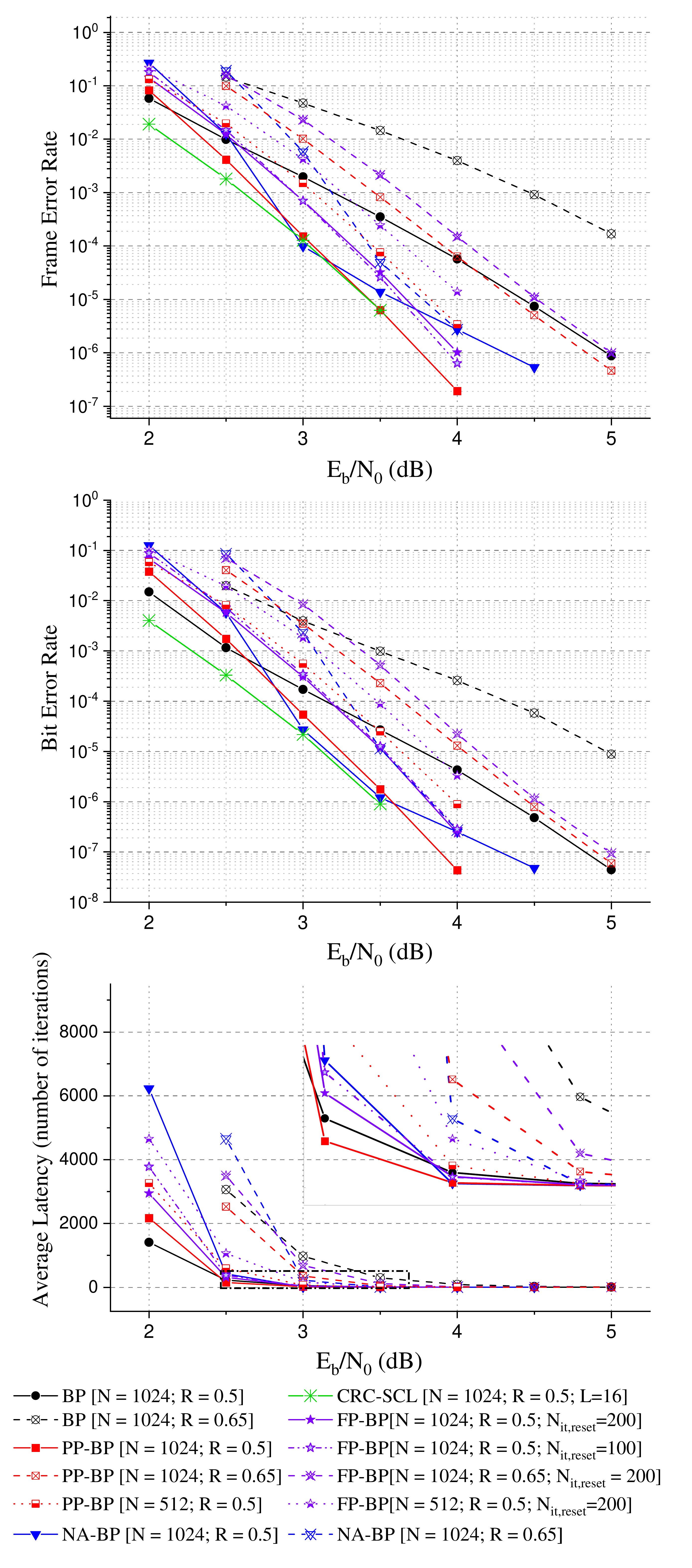}
  \caption{BER, FER and average latency comparison between BP, NA-BP, PP-BP, FP-BP and CRC-SCL for $N\!=\!512/1024$ and $R\!=\!0.5/0.65$ polar code with 24-bit CRC and $N_{it,max} \!=\! 2\!\!\times\!\!10^4$}
  \label{full_perf}
\end{figure}

Figure \ref{full_perf} compares the performance of proposed PP-BP with BP, FP-BP, and NA-BP for a polar code with $N\!=\!512 \text{ and }1024$, $R\!=\!0.5\text{ and }0.65$, and $N_{it,max} \!=\! 2\!\times\!10^4$. Further, the results are compared against CA-SCL decoder with list size (L) 16. The saturating behavior in BP decoding when $N_{it,max}$ is increased can be observed, since the gain is insignificant  when $N_{it,max}$ is increased from $200$ to $2\times10^{4}$. On the contrary, for $R\!=\!0.5 \text{ and } R=0.65$ PP-BP shows significant improvements up to $1dB$ at $10^{-6}$ when $N_{it,max}$ is increased. Furthermore, when $R\!=\!0.5$ PP-BP outperforms FP-BP [$N_{it,reset}=200$] \cite{permuteFactor} and FP-BP [$N_{it,reset}=100$] decoders by $0.25dB$ and $0.18dB$ respectively at $10^{-6}$. Thus, it can be concluded that partial permutations are more effective than full permutations, particularly when oscillation errors dominate the error distribution. In FP-BP, even though some performance improvements can be obtained in high SNR regime by decreasing $N_{it,reset}$, a performance degradation at low SNR regime can be observed. While more frequent permutations are effective in high SNR regime where effects of cycles and stopping sets are prominent, in low SNR regime, the number of iteration in a single permutation has to be sufficiently large to achieve a convergence from noisy data. Although NA-BP shows significant improvements for both $R\!=\!0.5 \text{ and } R=0.65$ when $N_{it,max}$ is increased, severe error floor can be observed around $3.5dB$ in the case of $R\!=\!0.5$ emphasizing the ineffectiveness of single trellis based decoders against oscillation errors. Nonetheless, for $R\!=\!0.65$, NA-BP outperforms both PP-BP and FP-BP by about $0.6dB$ at $10^{-5}$ in high SNR regime proving the limited effectiveness of multi-trellis decoders against false convergences. 

From the results in Figure \ref{full_perf}, it can be observed that the adverse effect on the latency in low SNR regime is less in the PP-BP compared to NA-BP and FP-BP. However, for the case of $R\!=\!0.65$, NA-BP outperforms the PP-BP in terms of latency as the effectiveness of partial permutations is limited against false convergences. In terms of hardware complexity, the proposed PP-BP is comparable with that of FP-BP since the same hardware implemented for permuting the factor graph in FP-BP can be utilized. The only calculation carried out in addition to FP-BP is the $N_{it,reset}$ calculation at every permutation which is fixed in the FP-BP. However, determining the decoder specific parameters in design phase is a challenging task since it was observed that the effect of improper parameter values is only visible at high SNR regime resulting in an error floor.

\section{conclusion}

In this paper, we proposed a new variant of the multi-trellis BP decoder based on permutation of random subgraphs of the factor graph, enabling the decoder to retain information of all the variable nodes except for the ones inside the permuted subgraph. This results in more permutations within fixed number of iterations compared to multi-trellis decoders based on full permutations. Experimental results show that the proposed decoder is more effective against oscillation errors compared to converged errors. For a polar code with $N\!=\!1024$ and $R=0.5$, $0.25dB$ gain at $10^{-6}$ compared to multi-trellis decoder with full permutations proposed in \cite{permuteFactor} can be observed along with a significant improvement to latency in low SNR regime. In a future step, the use of noise injection along with partial permutations to improve the effectiveness against converged errors needs to be investigated. 

 \IEEEpubidadjcol 

\renewcommand*{\bibfont}{\scriptsize}

\printbibliography

\end{document}